\documentclass[12pt]{iopart}
\usepackage{graphicx}

\newtheorem{definition}{Definition}
\usepackage{enumerate}

\begin{document}

\title[Personalized recommendation with corrected similarity]{Personalized recommendation with corrected similarity}

\author{Xuzhen Zhu$^1$, Hui Tian${^{1}}$\footnote[7]{Corresponding author}, Shimin Cai$^2$}
\address{$^1$State Key Laboratory of Networking and Switching Technology, Beijing University of Posts and Telecommunications, Beijing, 100876, P.R.China}
\address{$^2$Web Sciences Center, University of Electronic Science and Technology of China, Chengdu, 610054, P.R.China}
\ead{tianhui@bupt.edu.cn}
\begin{abstract}
Personalized recommendation attracts a surge of interdisciplinary researches.
Especially, similarity based methods in applications of real recommendation
systems achieve great success. However, the computations of similarities are
overestimated or underestimated outstandingly due to the defective strategy of
unidirectional similarity estimation. In this paper, we solve this
drawback by leveraging mutual correction of forward and backward similarity estimations,
and propose a new personalized recommendation index, i.e., corrected similarity
based inference (CSI). Through extensive experiments on four benchmark datasets,
the results show a greater improvement of CSI in comparison with these mainstream baselines.
And the detailed analysis is presented to unveil and understand the origin of such
difference between CSI and mainstream indices.
\end{abstract}

\maketitle

\section{Introduction}

With the revolutionary development of technology of Internet~\cite{zhang2008evolution,pastor2007evolution},
World Wide Web~\cite{broder2000graph,doan2011crowdsourcing} and smart mobile devices~\cite{goggin2012cell,zheng2010smart},
information bursts out explosively and inconceivably changes lifestyle of human beings~\cite{schafer1999recommender}.
Different from traditional lifestyle, people are gradually accustomed to acquiring information through an online way,
such as reading news in web portals, watching movies in video websites, shopping in E-commerce platform, etc.
While information is continuously growing day by day, billions of objects, involving with millions of movies,
songs and books, and numerous news, become overloaded and severely challenge personal processing abilities.
It leads people to an awkward and painful situation that they cannot find favorite objects due to their
limited searching abilities, in contrary, numerous objects are unknown to the needed people,
that is so-called long-tail effect~\cite{anderson2008long}.
Facing such situation, the personalized recommendation technology~\cite{resnick1997recommender}
comes out to break the dilemma, which captures peoples' habits through historical records of
visiting and browsing activities in websites and further based on the habits
recommend objects to the needed people. For examples, \textit{Amazon.com} uses
purchase records to recommend books~\cite{linden2003amazon},
\textit{AdaptiveInfo.com} uses reading histories to recommend news~\cite{billsus2007adaptive},
and \textit{TiVo} recommends TV shows and movies on
the basis of users' viewing patterns and ratings~\cite{ali2004tivo}.

Driven by the great significance in economy and society~\cite{huang2007comparison,wei2007survey},
a large quantity of studies on recommendation systems are ever-lastingly achieved in various fields, from science analysis
to engineering practice and from computer science to physics community (see the review articles~\cite{adomavicius2005toward,lu2012recommender}
and the references therein).
Fruitful personalized recommendation technologies~\cite{shapira2011recommender} are come up with and applied in real environments,
including content-based analysis~\cite{ansari2000internet,pazzani2007content}, knowledge-based analysis~\cite{trewin2000knowledge},
context-aware analysis~\cite{adomavicius2005incorporating}, time-aware analysis~\cite{petridou2008time,campos2013time},
tag-aware analysis~\cite{zhang2011tag,tso2008tag}, social recommendation analysis~\cite{liu2014new,guy2011social},
constraint-based analysis\cite{felfernig2008constraint}, spectral analysis~\cite{maslov2001extracting},
iterative refinement~\cite{ren2008information}, principle component analysis~\cite{goldberg2001eigentaste}, etc.
Besides, some similarity based recommendation algorithms due to high simplicity and effectiveness obtains
widespread applications in personalized recommendation systems, such as collaborative filtering~\cite{herlocker2004evaluating},
network based inference~\cite{zhou2007bipartite,zhou2009accurate,kleinberg2007cascading}, diffusion-based algorithms~\cite{zhang2007recommendation,zhou2008effect,pei2013spreading,kitsak2010identification},
and hybrid spreading \cite{burke2002hybrid,zhou2010solving}.

\begin{figure}
\begin{center}
\includegraphics[width=10cm]{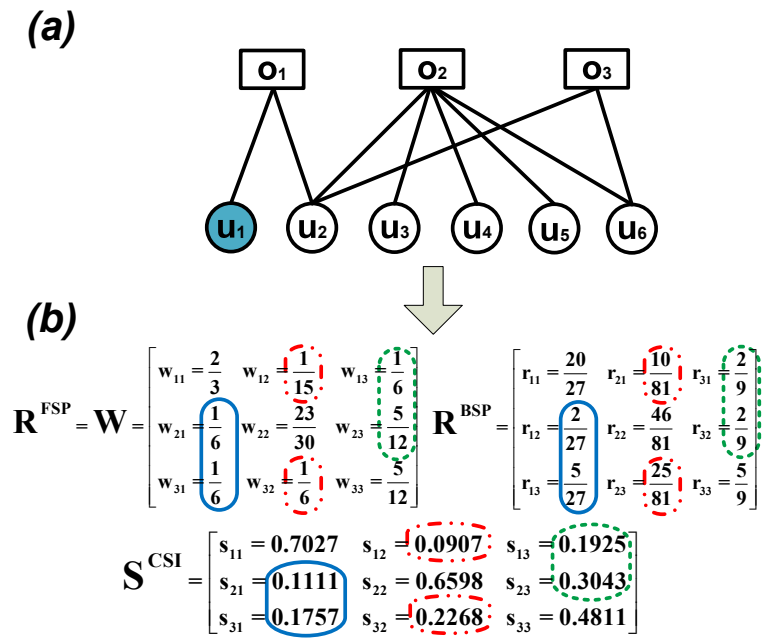}\\
\caption{Illustrating the correction of similarity. (a) the description of a unweighted
undirected object-user bipartite network, with objects denoted by squares
and users by round circles. (b) Matrices $W$, $R^{FSP}$, $R^{BSP}$ and $S^{CSI}$
indicate network based similarity matrix, forward and backward similarity
proportion matrices and corrected similarity matrix, respectively.
Color circles highlight the corresponding relations of similarity
elements in different similarity matrices. Here element $s_{ij}$ in $S^{CSI}$ equals to
$\sqrt{w_{ij}\times r_{ji}}$, with $w_{ij}$ in $W=R^{FSP}$ and $r_{ji}$ in $R^{BSP}$.}\label{fig:CSI}
\end{center}
\end{figure}

In an unweighted undirected object-user bipartite network (BN), the basic theory in these
similarity based methods has supposed that two objects are believed
to be similar if they are simultaneously selected by a user, and
the more users they are selected by, the more similar they are believed to be.
However, because of sparsity and complexity in BN, in fact, some similarities
among pairs of objects/users are overestimated or underestimated outstandingly,
which generates many fake similarities leading to a lower recommendation accuracy.
Here we take a specific example to explain the origin of problem,
Figure~\ref{fig:CSI}(a) describes a BN, with the same condition that
object $o_1$ and $o_2$, $o_1$ and $o_3$ are only selected by user $u_2$ at the same time, so that
the similarity from $o_1$ to $o_2$ is expected to the same as the one from $o_1$ to $o_3$.
Nevertheless, it deviates from this expectation, that is, the statistical sums of
similarities between each object and others are the same to be set as 1,
and bidirectional similarities are essentially the same. In total five users selecting $o_2$,
only one also selects $o_1$ and for $o_3$ it is one in two. For $o_2$, the most likely
similarity only accounts for $\frac{1}{5}$ of the original, and for $o_3$ it
accounts for $\frac{1}{2}$. Thus, it suggests that the
original similarity is overestimated between
$o_1$ and $o_2$ or underestimated between $o_1$ and $o_3$.
In further, the discussed difference of similarities implies the existing drawback of the basic
theory in these similarity based methods. To solve this drawback, a new method that can sensitively find and
represent this difference is urging to be well designed.

To fairly handle the drawback of existing similarity estimation, herein we leverage the forward and
backward similarity proportions to correct it, and according to
that propose a new personalized recommendation index named as corrected similarity based inference(CSI)
based on BN to enhance recommendation performances.
Through extensive experiments on four benchmark datasets (Movielens, Netflix, Amazon and RYM), the
results showing a great improvement of CSI in comparison of these mainstream baselines suggests its effectiveness.

The rests of paper are organized as follows: in section 2, the new model based on corrected similarity is introduced;
in section 3 and 4, the experimental materials of four benchmark datesets and methods including metrics and four mainstream baselines
are described respectively; we present the results and discussions in section 5 and finally make a conclusion.

\section{The corrected similarity based inference model}
A recommendation system commonly consists of users and objects,
in which each user has collected some objects. By denoting the object-set as $O = \{o_1, o_2, \cdots, o_n\}$
and user-set as $U = \{u_1, u_2, \cdots, u_m\}$, the recommendation system can be fully described by an $n \times m$
adjacent matrix $A = \{a_{ij}\}$, where $a_{ij} = 1$ if $o_i$ is collected by $u_j$,
and $a_{ij} = 0$ otherwise. Thus, a recommendation system can be also described as a BN $G(O,U)$.
The personalized recommendation index according to BN based similarity shows lower complexity, higher
effectiveness and more outstanding personality than traditional ones, achieving a lot of significant applications
and continuously attracts widespread attention~\cite{lu2012recommender}.
Here, we take BN based similarity to build our corrected similarity based inference model.

\subsection{Bipartite network based similarity}
First, we introduce the personalized recommendation index according to BN based similarity.
In Ref.~\cite{zhou2007bipartite}, it builds an object based relation network and defines
object-similarity weight between object $o_i$ and $o_j$ as below:
\begin{equation}\label{eq:NBI}
w_{ij} = \frac{1}{k(o_j)}\sum_{l=1}^{m}\frac{a_{il}a_{jl}}{k(u_l)}
\end{equation}
wherein $w_{ij}$ denotes the similarity between $o_i$ and $o_j$ and points out
that how much probability a user will be recommended $o_i$ if he/her has selected $o_j$.
Furthermore, if such user's selections can be represented by a vector $f$, the
coming recommendations are $f'=Wf$, with $W=\{w_{ij}\}$ denoted as the similarity matrix.

Then taking figure~\ref{fig:CSI} for example, $G$ contains user-set $O=\{o_1,o_2,o_3\}$ and user-set $U=\{u_1,u_2,u_3,u_4,u_5,u_6\}$,
and all selections are exhibited together. According to Equ.~(\ref{eq:NBI}), all similarity weights between pairs of objects are obtained in matrix
$W$ in Fig.\ref{fig:CSI}(b). Generally, a user wouldn't be recommended the objects he/she has selected, so we just emphasize the similarities
between two different objects via different circles. Belonging to similarity based recommendation, this network based recommendation index also suffers
the formerly discussed fake of similarity estimation which can be found in $w_{21}$ and $w_{31}$ holding the same weights surrounded by blue solid circle,
even though the similarities in other circles are fairly distinguishably estimated. Consequently, the decision cannot be made to recommend which one
of $o_2$ and $o_3$ to user $u_1$ marked blue in Fig.\ref{fig:CSI}(a). This dilemma is urgent to be solved by correcting the computation of similarity.

\subsection{Corrected similarity}
The reason resulting in the fake similarity is sparsity and asymmetrical
estimation that only considers the unidirectional similarity directly used in recommendation, such as from
$o_1$ to $o_2$ via $w_{21}$ of $W$ in Fig.~\ref{fig:CSI}(b). Much more practically, two objects are believed to be
similar only if the forward similarity proportion is coherent with the backward similarity
proportion. And the more coherent, the more similar they are. We give the definitions
of the forward and backward similarity proportions as follows:
\begin{definition}\label{def:FSP_BSP}
Given BN $G(O,U)$, similarity weight matrix $W=\{w_{ij}\}$ denotes the similarity between
$o_i$ and $o_j$. The element $r^{FSP}_{ij}$ of forward similarity proportion matrix $R^{FSP}$ can defined by
the ratio between $w_{ij}$ and $\sum_{i=1}^{n}w_{ij}$ and can be delivered as below, since $\sum_{i=1}^{n}w_{ij}=1$.
\begin{equation}
r^{FSP}_{ij} = \frac{w_{ij}}{\sum_{i=1}^{n}w_{ij}} = w_{ij}
\end{equation}
likewise, backward similarity proportion matrix $R^{BSP}=\{r^{BSP}_{ji}\}$ can be define as follows:
\begin{equation}
r^{BSP}_{ji} = \frac{w_{ji}}{\sum_{j=1}^{n}w_{ji}} = r_{ji},
\end{equation}
where $r^{BSP}_{ji}$  is simplified as $r_{ji}$.
\end{definition}
\begin{definition}
Then, based on  $r^{FSP}_{ij}$ and  $r^{BSP}_{ji}$, the corrected similarity $s^{CSI}_{ij}$ can be defined as:
\begin{equation}
s^{CSI}_{ij} = \sqrt{r^{FSP}_{ij}\times r^{BSP}_{ji}},
\end{equation}
where the similarity can be comprehensively corrected by forward similarity proportion
$r^{FSP}_{ij}$ and backward similarity proportion $r^{BSP}_{ji}$ at the same time
and the greater the corrected similarity is, the more similar the two objects identically are.
\end{definition}

If an user has selections denoted by vector $f$, the recommendations $f'$ using corrected
similarity matrix $S^{CSI}$ can be derived from the equation $f'=S^{CSI}f$.
In Fig~\ref{fig:CSI}(b), $R^{FSP}$ and $R^{BSP}$ are illustrated. The original fake similarity
estimations of $w_{21}$ and $w_{31}$ in blue solid circle, through the corrections
via $r_{12}$, $r_{13}$ and definition of $r^{BSP}_{ji}$, are corrected as $s_{21}$ and
$s_{31}$, between which the clear difference is embodied and confirms our formally
expectation. Meanwhile, other similarity weight $w_{ij}$ are transformed into $s_{ij}$
with the same circle marker, keeping the existing distinguishability, such
as $w_{13}$ and $w_{23}$ into $s_{13}$ and $s_{23}$ surrounded by green dash circles.

\section{Experimental data}
\renewcommand*{\thefootnote}{\alph{footnote}}
\setcounter{footnote}{0}
For demonstrating the excellent effectiveness and efficiency of CSI, we introduce four real benchmark datasets,
\textit{Movielens}\footnote{http://www.grouplens.org/},
\textit{Netflix}\footnote{http://www.netflix.com/},
\textit{Amazon}\footnote{http://www.amazon.com/} and
\textit{RYM}\footnote{http://rateyourmusic.com/}, as experimental materials
the first two are from famous movie recommendation websites, the third is from a well-known online shopping store, and the last is from a music recommendation website.
To recommend the appropriate objects, they all leverage ratings to capture users' preferences, with rating from 1 to 5 stars in \textit{Movielens, Netflix} and \textit{Amazon} and from 1 to 10 in \textit{RYM}.
User is believed to like the object as a user-object link, if the ratings $\ge 3$ in \textit{Movielens}, \textit{Netflix}, \textit{Amazon} and $\ge5$ in \textit{RYM}.
After deleting the `dislike' links, we obtain the experimental datasets with detailed information in Tab.~\ref{tab:dataset}.\\
\begin{table}[thp]
\caption{Summary on primary information of four datasets}
\label{tab:dataset}
\setlength{\tabcolsep}{5pt}
\begin{center}
\begin{tabular}{ccccc}
\hline\hline
Data  & Users & Objects & Links & Sparsity\\
\hline
Movielens& 943 & 1682 & 1000000 & $6.3\times10^{-1}$\\
Netflix& 10000 & 6000 & 701947 & $1.17\times10^{-2}$\\
Amazon& 3604 & 4000 & 134679 & $9.24\times10^{-3}$\\
RYM& 33786 & 5381 & 613387 & $3.37\times10^{-3}$\\
\hline\hline
\end{tabular}
\end{center}
\end{table}

In the experiments, all the possible user-object links constitute a total link-set $E^A$.
The existed link-set $E$ should be divided into, training set $E^T$ including 90$\%$ links of the total
and testing set $E^P$ containing the rest 10$\%$ links, with $E^P \setminus E^T=\emptyset$, obviously. Notice that
the links in testing set are regarded as unknown information and forbidden from using in training process.
The difference between $E^A \setminus E$ contains all the ultimately unrealized user-object links.

\section{Experimental methods}

\subsection{Metrics}
A personalized recommendation index is always focused on three classes of performances: accuracy, diversity and popularity~\cite{lu2012recommender}.
The accuracy is usually assessed by three metrics, including averaged ranking score, precision and AUC, which are described as follows:
\begin{enumerate}[(1)]
\item Averaged ranking score ($\langle r\rangle$):
Ranking score evaluates the extent the entire user-object links in the testing set $E^P$ are ranked ahead to in the user-object link set $E^A \setminus E^T$. If $o_i$ is selected by $u_j$ in the $E^P$ and has the position $p_{ij}$ in $u_j$'s uncollected objects set $O_j$ according to the recommendation score, we have $rank_{ij}=\frac{p_{ij}}{\left|O_j\right|}$ as the ranking score of $o_i$-$u_j$ link $l_{ij}$. Eventually, the averaged ranking score $\langle r\rangle$ over all the links in $E^P$ equals to:
\begin{equation}
\langle r\rangle = \frac{\sum_{l_{ij}\in E^P}rank_{ij}}{\left|E^P\right|}
\end{equation}
Where $\left|O_j\right|$ and $\left|E^P\right|$ all indicate the number of elements in a set.
\item Precision ($P$): Precision measures the ratio in which how many links in the $E^P$ are eventually selected in every user's recommendation list with length $L$, so the precision $P_j(L)$ of user $u_j$ equals to $\frac{N_j}{L}$ with $N_j$ standing for the number of the recommended testing links. Therefore, the definition of the whole system, averaged on individual precisions over all users, looks as follows:
\begin{equation}\label{ep:P-L}
P = \frac{1}{m}\sum_{j=1}^{m}P_j(L)
\end{equation}
\item AUC: AUC (Area Under ROC Curve) attempts to measure how a recommender system can
successfully distinguish the relevant objects (those appreciated by a user) from the irrelevant objects (all the others). The
simplest way to calculate AUC is by comparing the probability that the relevant objects will be recommended with that of
the irrelevant objects. For n independent comparisons (each comparison refers to choosing one relevant and one irrelevant
object), if there are $n'$ times when the relevant object has higher score than the irrelevant and $n''$ times when the scores are equal, then
\begin{equation}
AUC=\frac{n'+0.5n''}{n}
\end{equation}
Clearly, if all relevant objects have higher score than irrelevant objects, AUC = 1 which means a perfect recommendation
list. For a randomly ranked recommendation list, AUC = 0.5. Therefore, the degree of which AUC exceeds 0.5 indicates
the ability of a recommendation algorithm to identify relevant objects.
\end{enumerate}

The diversity is usually evaluated by intra-similarity and hamming distance, of which the details are shown as below:
\begin{enumerate}[(1)]
\item Intra-similarity ($I$):
A good algorithm should also make the recommendations to a single user diverse to some
extent \cite{ziegler2005improving}, otherwise users may feel tired for receiving many recommended objects
under the same topic. Therefore, for an arbitrary target user $u_l$,
denoting the recommended objects for $u_l$ as \{$o_1$,$o_2$,..., $o_L$\}, Using also the S$\phi$ensen index\cite{sorensen1948method},
the similarity between two objects, $o_i$ and $o_j$ , can be written as:
\begin{equation}\label{ep:sij-o}
s_{ij}^{o}=\frac{1}{\sqrt{k(o_i)k(o_j)}}\sum_{l=1}^{m}a_{il}a_{jl}
\end{equation}
$k(o_i)$ means the degree of item $i$. The intra-similarity of $u_l$'s recommendation list can be defined as:
\begin{equation}\label{ep:intra-simi-l}
I_l=\frac{1}{L(L-1)}\sum_{i\ne j}s_{ij}^{n}
\end{equation}
The intra-similarity of the whole system is thus defined as:
\begin{equation}\label{ep:intra-simi}
I=\frac{1}{m}\sum_{l=1}^{m}I_l
\end{equation}

\item Hamming distance ($H$):
The algorithm should guarantee the diversity of recommendations, viz., different users should be recommended
different objects. It is also the soul of personalizedized recommendations. The intra-diversity
can be quantified via the \textit{Hamming distance} \cite{zhou2008effect}. Denoting $L$ the length
of recommendation list (i.e., the number of objects recommended to each user), if the
overlapped number of objects in $u_i$ and $u_j$'s recommendation lists is $Q$, their Hamming distance is defined as:
\begin{equation}\label{ep:Hamming}
H_{ij} = 1 - Q/L
\end{equation}
Generally speaking, a more personalized recommendation list should have larger Hamming distances to other lists.
Accordingly, we use the mean value of Hamming distance,
\begin{equation}\label{ep:mean-Hamming}
H = \frac{1}{m(m-1)}\sum_{i\ne j}H_{ij}
\end{equation}
averaged over all the user-user pairs, to measure the diversity of recommendations.
Note that, $H$ only takes into account the diversity among users.
\end{enumerate}

The popularity is closely related to personality, and estimated by average degree over recommended objects:
\begin{enumerate}[(1)]
\item Average degree ($\langle k\rangle$):
Given $o_{ij}$ is the j$th$ recommended item for user $i$, $k(o_{ij})$ represents the degree of item $o_{ij}$, so the popularity is defined as the average degree of all recommended items for all users as follows:
\begin{equation}\label{ep:popularity}
<k>=\frac{1}{mL}\sum_{i=1}^{m}\sum_{j=1}^{L}k(o_{ij})
\end{equation}
\end{enumerate}

\subsection{Baselines}
For demonstrating the greater improvement compared with classical
similarity based methods, four mainstream indices,
global ranking method(GRM), cooperative filtering (CF),
network based inference (NBI), initial configuration of NBI(IC-NBI),
are introduced below:
\begin{enumerate}[(1)]
\item GRM~\cite{herlocker2004evaluating}:
Supposed the user-set is \{$u_1$, $u_2$,...,$u_m$\}, the object-set is \{$o_1$, $o_2$, ..., $o_n$\}
and the degree-set of all items is \{k($o_1$), k($o_2$), ..., k($o_n$)\}.
In GRM, it sorts all the objects in the descending order of degree and recommends those with the highest degrees,
i.e., we list the degrees in the descending order as k($o_{i1}$) $\ge$ k($o_{i2}$) $\ge$  ...  $\ge$ k($o_{in}$).
At last, after eliminating the objects that have collected from the descending order, the top-$L$ objects in the rest are the recommended items.

\item CF~\cite{herlocker2004evaluating}:
Collaborative filtering measures the similarity between users or objects. For two users $u_i$ and $u_j$, their cosine similarity is defined as (for more local similarity indices as well as the comparison of them, see the Refs. \cite{liben2007link,zhou2009predicting}):
\begin{equation}\label{ep:UCF-simi}
s_{ij}=\frac{1}{\sqrt{k(u_i)k(u_j)}}\sum_{l=1}^{n}a_{li}a_{lj}
\end{equation}
For any user-object pair $u_i-o_j$ , if $u_i$ has not yet collected $o_j$ (i.e., $a_{ji}$ = 0), the predicted
score, $v_{ij}$ (to what extent $u_i$ likes $o_j$), is given as
\begin{equation}\label{ep:UCF-rec}
v_{ij}=\frac{\sum_{l=1,l\ne i}^{m}s_{li}a_{jl}}{\sum_{l=1,l\ne i}^{m}s_{li}}
\end{equation}
For any user $u_i$, all the nonzero $v_{ij}$ with $a_{ji} = 0$ are sorted in a descending order, and those objects in the top-$L$ are recommended.
\item NBI~\cite{zhou2007bipartite}:
NBI is an algorithm based on network structure, and also uses the the S$\phi$ensen index. For a general user-object network, the similarity weight between $o_i$ and $o_j$ reads:
\begin{equation}\label{ep:NBI-w}
w_{ij}^{NBI}=\frac{1}{k(o_j)}\sum_{l=1}^{m}\frac{a_{il}a_{jl}}{k(u_l)}
\end{equation}
where $w_{ij}^{NBI}$ belongs to similarity weight matrix $W^{NBI}$, and $k(o_j) = \sum_{i=1}^{m}a_{ji}$ and $k(u_l) = \sum_{i=1}^{n} a_{il}$ respectively denote the degrees of object $o_j$ and user $u_l$.
The recommendation list of user $u_l$ is $f_l'=W^{NBI}f_l$, with ${f_l}={a_{li}}$ representing the historical record of $u_l$.
\item IC-NBI~~\cite{zhou2008effect}:
IC-NBI is a modified NBI algorithm dependent on initial resource configuration with weight $w_{ij}^{IC-NBI}=k(o_j)w_{ij}$. $w_{ij}$ is referred in Eq.~(\ref{ep:NBI-w}) and $W^{IC-NBI}=\{w_{ij}^{IC-NBI}\}$. With selection history $f_j$ of $u_j$, the recommendation list of $u_j$ is $f_j'=W^{IC-NBI}f_j$.

\end{enumerate}
\begin{table}[tp]
\begin{center}
\caption{Performance comparison table.
The $\langle r\rangle$ for ranking score, $P$ for precision and $AUC$ of IC-NBI are adopted at the optimal $\beta$ of each metric, and other metrics---$I$ for intra-similarity, $H$ for hamming distance, $\langle k\rangle$ for popularity---take the values corresponding to the optimal $\beta$ of $\langle r\rangle$.
The recommendation list $L=50$, and the sampling number $n$ in $AUC$ is one million. All the values are obtained by averaging
over ten independent runs with different data set divisions and numbers in brackets stand for the standard deviations.}
\label{tab:performances}
\setlength{\tabcolsep}{2pt}
\scriptsize{
\begin{tabular}{ccccccc}
\hline\hline
Movielens  & $\langle r\rangle$ & $P$ & $AUC$ & $I$ & $H$ & $\langle k\rangle$\\
\hline
GRM& 0.1486(0.0020) & 0.0508(0.0007) & 0.8569(0.0023) & 0.4085(0.0010) & 0.3991(0.0007) & 259(0.4410) \\
CF& 0.1225(0.0020) & 0.0638(0.0011) & 0.8990(0.0020) & 0.3758(0.0008) & 0.5796(0.0016) & 242(0.3724) \\
NBI& 0.1142(0.0018) & 0.0670(0.0011) & 0.9093(0.0016) & 0.3554(0.0008) & 0.6185(0.0013) & 234(0.3925) \\
IC-NBI& 0.1074(0.0017) & 0.0693(0.0011) & 0.9145(0.0014) & 0.3392(0.0009) & 0.6886(0.0011) & 219(0.4725) \\
CSI&\textbf{0.0963(0.0014)} & \textbf{0.0738(0.0009)} & \textbf{0.9276(0.0012)} & \textbf{0.2892(0.0008)}&\textbf{0.7601(0.0006)}&\textbf{186(0.4286)}\\
\hline
Netflix  & $\langle r\rangle$ & $P$ & $AUC$ & $I$ & $H$ & $\langle k\rangle$\\
\hline
GRM&  0.2046(0.0004) & 0.0160(0.0002) & 0.8101(0.0028) & 0.3580(0.0021) & 0.1627(0.0004) & 520(1.3402)\\
CF& 0.1755(0.0004) & 0.0235(0.0003) & 0.8714(0.0021) & 0.3106(0.0009) & 0.6787(0.0010) & 423(1.2803)\\
NBI& 0.1661(0.0004) & 0.0251(0.0003) & 0.8858(0.0019) & 0.2819(0.0008) & 0.7299(0.0006) & 398(1.0763)\\
IC-NBI& 0.1537(0.0004) & 0.0270(0.0004) & 0.8877(0.0020) & 0.2405(0.0006) & 0.8790(0.0003) & 312(0.6855)\\
CSI& \textbf{0.1437(0.0003)} & \textbf{0.0310(0.0004)} & \textbf{0.9063(0.0016)} & \textbf{0.1937(0.0012)}&\textbf{0.9063(0.0003)}&\textbf{256(0.7554)}\\
\hline
Amazon  & $\langle r\rangle$ & $P$ & $AUC$ & $I$ & $H$ & $\langle k\rangle$\\
\hline
GRM&0.3643(0.0017) & 0.0036(0.00008) & 0.6409(0.0029) & \textbf{0.0709(0.0006)} & 0.0584(0.0001) & 133(0.3) \\
CF&0.1212(0.0010) & 0.0156(0.0001) & 0.8810(0.0017) & 0.0927(0.0001) & 0.8649(0.0008) & 81(0.1938) \\
NBI&0.1169(0.0011) & 0.0161(0.0001) & 0.8844(0.0018) & 0.0899(0.0001) & 0.8619(0.0006) & 81(0.1775)\\
IC-NBI&0.1169(0.0014) & 0.0163(0.0001) & 0.8844(0.0018) & 0.0896(0.0001) & 0.8652(0.0006) & 81(0.1689)\\
CSI&\textbf{0.1036(0.0011)} & \textbf{0.0190(0.0001)} & \textbf{0.8930(0.0018)} & 0.0880(0.0002)&\textbf{0.9667(0.00007)}&\textbf{48(0.0479)}\\
\hline
RYM  & $\langle r\rangle$ & $P$ & $AUC$ & $I$ & $H$ & $\langle k\rangle$\\
\hline
GRM& 0.1581(0.00009) & 0.0034(0.00001) & 0.8786(0.0001) & \textbf{0.1334(0.0003)} & 0.0701(0.00007) & 1343(0.4268)\\
CF& 0.0753(0.0001) & 0.0129(0.00003) & 0.9548(0.0001) & 0.1604(0.00006) & 0.8216(0.00001) & 1114(0.5895)\\
NBI& 0.0673(0.00007) & 0.0131(0.00006) & 0.9611(0.0001) & 0.1580(0.0001) & 0.7912(0.00008) & 1195(0.7061)\\
IC-NBI& 0.0587(0.00007) & 0.0135(0.00005) & 0.9644(0.0001) & 0.1548(0.00008) & 0.8113(0.00001) & 1154(0.5654)\\
CSI&\textbf{0.0462(0.0001)} & \textbf{0.0156(0.00003)} & \textbf{0.9714(0.0001)} & 0.1467(0.00009)&\textbf{0.8922(0.00005)}&\textbf{869(0.5121)}\\
\hline\hline
\end{tabular}
}
\end{center}
\end{table}

\section{Results and discussions}
\begin{figure}[t]
\begin{center}
  \includegraphics[width=11cm]{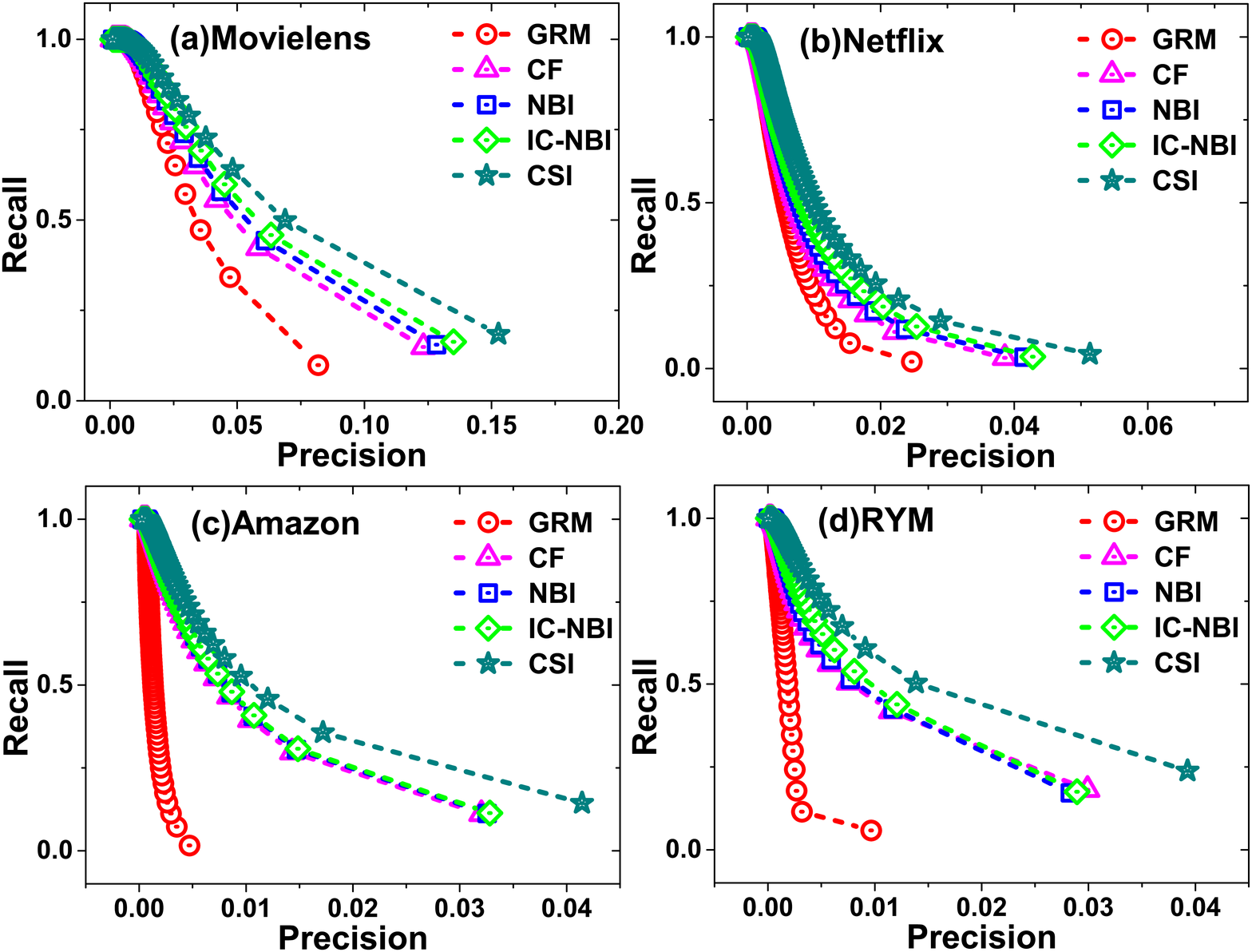}\\
  \caption{Demonstration of precision-recall curves on four datasets. Given a recommendation list length $L$, a precision and recall can be respectively achieved via Equ.(\ref{ep:P-L}) and $\frac{l}{\left|E^P\right|}$, with $l$ and $\left|E^P\right|$ denoting the number of all hitting links in testing set and the size of testing set, respectively. When $L$ varies from 1 to the size of testing set $\left|E^P\right|$, we obtain the whole precision-recall curve (referenced in \cite{powers2011evaluation}). GRM, CF, NBI, IC-NBI and CSI are identically in four subgraphs plotted from the bottom left to upper right. CSI outstandingly outperforms others.  }\label{fig:pr}
\end{center}
\end{figure}

The experiments results on the above-mentioned four benchmark datasets are averaged over ten independent random divisions.
For convenient exhibition of differences, table~\ref{tab:performances} organizes all related performance indices, and
furthermore figure~\ref{fig:pr} depicts the precision-recall curves on four datasets for more intuitively demonstrating the
performance of performance.


As shown in Tab.~\ref{tab:performances}, the optimal values of each index on six metrics are presented, and we can clearly find that the
best ones emphasized in boldface are almost obtained through CSI. Concretely speaking,
CSI surpasses GRM the most in all aspects, especially even with $\langle r\rangle$ reduced by
more than 71\% in Movielens, $P$ increased by more than 4 times, $H$ increased by more than 15 times
and $\langle k\rangle$ reduced by more than 63$\%$ in Amazon.
Although being better than GRM, CF is still worse than CSI in all metrics, and outstandingly,
CSI exceeds it overwhelmingly with $\langle r\rangle$ reduced by more than 38$\%$ in RYM, $P$
increased by more than 31$\%$, $I$ reduced by more than 37$\%$ and $H$ reduced by 33$\%$
in Netflix, and $\langle k\rangle$ reduced by more than 40$\%$ in Amazon.
Obtained further more improvement than CF, NBI is still defeated by CSI.
CSI transcends NBI on six metrics, distinctively, with $\langle r\rangle$
reduced by more than 31$\%$ in RYM, $P$ increased by more than 23$\%$, $I$ reduced by 31$\%$
and $H$ increased by more than 24$\%$ in Netflix, and $\langle k\rangle$ reduced by more than 40$\%$
in Amazon.
At last, IC-NBI considering more factors is the best in all baselines, but CSI still stands on top of it, remarkably,
with $\langle r\rangle$ reduced by more than 21$\%$ in RYM, $P$ increased by more than 16$\%$, $H$ increased by more than 11$\%$,
$\langle k\rangle$ reduced by more than 40$\%$ in Amazon and $I$ reduced by more than 19$\%$ in Netflix.
From statistical analysis of results in Tab.~\ref{tab:performances},
we argue that CSI obviously outperforms the four mainstream baselines in accuracy, diversity and personality,
even though there exist different degrees of improvement from GRM to IC-NBI. Especially, CSI acquires considerable improvement
in contrast to NBI according to the corrected similarity theory illustrated in Fig.~\ref{fig:CSI}.

The corresponding precision-recall curves of CSI and baselines on four datasets are plotted in Fig.~\ref{fig:pr}.
Given a recommendation list length $L$, a precision via Equ.(\ref{ep:P-L}) and recall via
$\frac{l}{\left|E^P\right|}$ can be obtained, respectively. Note that $l$ and $\left|E^P\right|$ denotes
the number of all hitting links in testing set and the size of testing set, respectively.
When $L$ varies from 1 to the size of testing set $\left|E^P\right|$, we achieve the whole
precision-recall curve (referenced in \cite{powers2011evaluation}).
According to the above method, in Fig.~\ref{fig:pr} labeling $x$-axis as precision and $y$-axis as recall,
precision-recall curves on four datasets are plotted, of which the identically descending order from
the bottom left to the upper right in four subgraphs suggests that the great difference of performance between CSI
and baselines and CSI absolutely outperforms all baselines. It further confirm the statistical results in Tab.~\ref{tab:performances}.

To unveil the underlying origin of more considerable performance improvement of CSI than baselines, we compare
the recommendation processes of these similarity based methods. Generally, GRM tends to
recommend the most popular objects to the user with the poorest similarity consideration,
undoubtedly leading to the worst performance of all mentioned metrics; CF
reasonably based on similarity between users obviously improves performances on six metrics but
still ranks with the second worst compared with CSI because of neglecting the similarity between
objects and unidirectional uncorrected similarity; NBI, based on network-projected object-similarity,
distinctively performs  better than CF but also shows severe shortage in contrast to CSI,
primarily still due to the unidirectional defective similarity representation between two objects;
IC-NBI considers not only the similarity between objects but also penalization on the high degree of popular objects
to further improve the performances, but inspite of complementary consideration of degrees of objects and inheriting
the uncorrected unidirectional defective similarity representation from NBI. In a word,
these traditional similarity based algorithms indeed contains the similar drawback of similarity estimation,
while CSI simultaneously considers the forward and backward similarity proportions to correct the
originally fake similarity estimations, surely achieving the considerable
improvements in accuracy, diversity and personality.

Besides, the low computation complexity is another important concern in the design of prediction
algorithm. As we known, the time complexity of product of two $N\times N$ matrices is $O(N^3)$. From the definitions of CF, NBI, IC-NBI, their time complexities are all $O(N^3)$. In contrast, although with same time complexity of $O(N^3)$, our index shows stronger performances than them. On the contrary, CF, using sorting algorithm with less time complexity of $O(N^2)$ approximately, exhibits the extremely worst performances. Above all, our index achieves best performance with no increase in complexity.
\section{Conclusions}
In conclusion, we have studied the similarity based recommendation algorithms (mainly involving with
baselines) and find that there are fake similarity estimations including underestimation and overestimation in them
due to only considering unidirectional similarity representation for recommendations.
After investigating the relation between forward and backward similarity estimations,
a corrected similarity based inference model (i.e., CSI) is proposed to make up the drawback
of traditional similarity based ones.
Through experimental verifications on four representative real datasets, CSI
indeed achieves great and impressive improvement in accuracy, diversity and personality (e.g., RYM),
compared with baselines.
Because of high effectiveness and low complexity, CSI can be applied in various kinds of recommendation environments,
such as online news recommendation, online books recommendation, online movies recommendation, online songs recommendation, and so on.
Although obtained great improvement, CSI still has weaknesses, for example, the lack of consideration on node degrees
which to some extent impacts the effectiveness of personalized recommendations.
In future, we will continue our research to further enhance the performances of personalized recommendation.
\section*{Acknowledgments}
This work was supported by National Major Science and Technology Special Project of China (No. 2012ZX03005010-003), National Natural Science Foundation of China (No. 61231009), National High Technology Research and Development Program of China (863 Program)(No. 2014AA01A706), and Funds for Creative Research Groups of China (No. 61121001).

\section*{References}

\begin{thebibliography}{10}
\expandafter\ifx\csname url\endcsname\relax
  \def\url#1{{\tt #1}}\fi
\expandafter\ifx\csname urlprefix\endcsname\relax\def\urlprefix{URL }\fi
\providecommand{\eprint}[2][]{\url{#2}}

\bibitem{zhang2008evolution}
Zhang G~Q, Zhang G~Q, Yang Q~F, Cheng S~Q and Zhou T 2008 {\em New J. Phys.\/}
  {\bf 10} 123027

\bibitem{pastor2007evolution}
Pastor-Satorras R and Vespignani A 2007 {\em Evolution and structure of the
  Internet: A statistical physics approach\/} (Cambridge University Press)

\bibitem{broder2000graph}
Broder A, Kumar R, Maghoul F, Raghavan P, Rajagopalan S, Stata R, Tomkins A and
  Wiener J 2000 {\em Comput. netw.\/} {\bf 33} 309

\bibitem{doan2011crowdsourcing}
Doan A, Ramakrishnan R and Halevy A~Y 2011 {\em Comm. ACM\/} {\bf 54} 86

\bibitem{goggin2012cell}
Goggin G 2012 {\em Cell phone culture: mobile technology in everyday life\/}
  (Routledge)

\bibitem{zheng2010smart}
Zheng P and Ni L 2010 {\em Smart phone and next generation mobile computing\/}
  (Morgan Kaufmann)

\bibitem{schafer1999recommender}
Schafer J~B, Konstan J and Riedi J 1999 Recommender systems in e-commerce {\em
  Proceedings of the 1st ACM Conf. Electro. Commer.\/} (ACM) p 158

\bibitem{anderson2008long}
Anderson C 2008 {\em The long tail: Why the future of business is selling less
  of more\/} (Hyperion Books)

\bibitem{resnick1997recommender}
Resnick P and Varian H~R 1997 {\em Comm. ACM\/} {\bf 40} 56

\bibitem{linden2003amazon}
Linden G, Smith B and York J 2003 {\em IEEE Int. Comput.\/} {\bf 7} 76

\bibitem{billsus2007adaptive}
Billsus D and Pazzani M~J 2007 {\em Adaptive news access\/} (Springer)

\bibitem{ali2004tivo}
Ali K and Van~Stam W 2004 Tivo: making show recommendations using a distributed
  collaborative filtering architecture {\em Proceedings of the 10th ACM SIGKDD
  Int. Conf.\/} (ACM) p 394

\bibitem{huang2007comparison}
Huang Z, Zeng D and Chen H 2007 {\em IEEE Intel. Syst.\/} {\bf 22} 68

\bibitem{wei2007survey}
Wei K, Huang J and Fu S 2007 A survey of e-commerce recommender systems {\em
  Proceedings of the Service Systems and Service Management Conference\/}
  (IEEE) p~1

\bibitem{adomavicius2005toward}
Adomavicius G and Tuzhilin A 2005 {\em IEEE Trans. Knowl. Data Eng.\/} {\bf 17} 734

\bibitem{lu2012recommender}
L{\"u} L, Medo M, Yeung C~H, Zhang Y~C, Zhang Z~K and Zhou T 2012 {\em Phys.
  Rep.\/} {\bf 519} 1

\bibitem{shapira2011recommender}
Shapira B 2011 {\em Recommender systems handbook\/} (Springer)

\bibitem{ansari2000internet}
Ansari A, Essegaier S and Kohli R 2000 {\em J. Market. Res.\/} {\bf 37} 363

\bibitem{pazzani2007content}
Pazzani M~J and Billsus D 2007 {\em Content-based recommendation systems\/}
  (Springer)

\bibitem{trewin2000knowledge}
Trewin S 2000 {\em Encyclopedia Libr. Inf. Sci.\/} {\bf 69} 69

\bibitem{adomavicius2005incorporating}
Adomavicius G, Sankaranarayanan R, Sen S and Tuzhilin A 2005 {\em ACM Trans.
  Inf. Syst.\/} {\bf 23} 103

\bibitem{petridou2008time}
Petridou S~G, Koutsonikola V~A, Vakali A~I and Papadimitriou G~I 2008 {\em IEEE
  Trans. Knowl. Data Eng.\/} {\bf 20} 653

\bibitem{campos2013time}
Campos P~G, D{\'\i}ez F and Cantador I 2013 {\em User Model User-Adap Inter.\/}
  {\bf 24} 67

\bibitem{zhang2011tag}
Zhang Z~K, Zhou T and Zhang Y~C 2011 {\em J. Comput. Sci. Technol.\/} {\bf 26}
  767

\bibitem{tso2008tag}
Tso-Sutter K~H, Marinho L~B and Schmidt-Thieme L 2008 Tag-aware recommender
  systems by fusion of collaborative filtering algorithms {\em Proceedings of
  the 2008 ACM Int. Symp. Comput.\/} (ACM) p 1995

\bibitem{liu2014new}
Liu H, Hu Z, Mian A, Tian H and Zhu X 2014 {\em Knowl-Based Syst.\/} {\bf 56}
  156

\bibitem{guy2011social}
Guy I and Carmel D 2011 Social recommender systems {\em Proceedings of the 20th
  Int. Conf. companion on World wide web\/} (ACM) p 283

\bibitem{felfernig2008constraint}
Felfernig A and Burke R 2008 Constraint-based recommender systems: technologies
  and research issues {\em Proceedings of the 10th Int. Conf. on electronic
  commerce\/} (ACM) p~3

\bibitem{maslov2001extracting}
Maslov S and Zhang Y~C 2001 {\em Phys. Rev. Lett.\/} {\bf 87} 248701

\bibitem{ren2008information}
Ren J, Zhou T and Zhang Y~C 2008 {\em EPL\/} {\bf 82} 58007

\bibitem{goldberg2001eigentaste}
Goldberg K, Roeder T, Gupta D and Perkins C 2001 {\em Inf. Retrieval\/} {\bf 4}
  133

\bibitem{herlocker2004evaluating}
Herlocker J~L, Konstan J~A, Terveen L~G and Riedl J~T 2004 {\em ACM Trans. Inf.
  Syst.\/} {\bf 22} 5

\bibitem{zhou2007bipartite}
Zhou T, Ren J, Medo M and Zhang Y~C 2007 {\em Phys. Rev. E\/} {\bf 76} 046115

\bibitem{zhou2009accurate}
Zhou T, Su R~Q, Liu R~R, Jiang L~L, Wang B~H and Zhang Y~C 2009 {\em New J.
  Phys.\/} {\bf 11} 123008

\bibitem{kleinberg2007cascading}
Kleinberg J 2007 {\em Algorithmic Game Theory\/} {\bf 24} 613

\bibitem{zhang2007recommendation}
Zhang Y~C, Medo M, Ren J, Zhou T, Li T and Yang F 2007 {\em EPL\/} {\bf 80}
  68003

\bibitem{zhou2008effect}
Zhou T, Jiang L~L, Su R~Q and Zhang Y~C 2008 {\em EPL\/} {\bf 81} 58004

\bibitem{pei2013spreading}
Pei S and Makse H~A 2013 {\em J. Stat. Mech.: Theory Exp.\/} {\bf 2013} P12002

\bibitem{kitsak2010identification}
Kitsak M, Gallos L~K, Havlin S, Liljeros F, Muchnik L, Stanley H~E and Makse
  H~A 2010 {\em Nat. Phys.\/} {\bf 6} 888

\bibitem{burke2002hybrid}
Burke R 2002 {\em User Model. User-Adap. Inter.\/} {\bf 12} 331

\bibitem{zhou2010solving}
Zhou T, Kuscsik Z, Liu J~G, Medo M, Wakeling J~R and Zhang Y~C 2010 {\em Proc.
  Nati. Acad. Sci.\/} {\bf 107} 4511

\bibitem{ziegler2005improving}
Ziegler C~N, McNee S~M, Konstan J~A and Lausen G 2005 Improving recommendation
  lists through topic diversification {\em Proceedings of the 14th Int. Conf.
  on World Wide Web\/} (ACM) p~22

\bibitem{sorensen1948method}
S{\o}rensen T 1948 {\em Biol. Skr.\/} {\bf 5} 1

\bibitem{liben2007link}
Liben-Nowell D and Kleinberg J 2007 {\em J. Am. Soc. Inf. Sci.\/} {\bf 58} 1019

\bibitem{zhou2009predicting}
Zhou T, L{\"u} L and Zhang Y~C 2009 {\em Eur. Phys. J. B\/} {\bf 71} 623

\bibitem{powers2011evaluation}
Powers D~M 2011 {\em J. Mach. Learn. Technol.\/} {\bf 2} 37

\end{thebibliography}

\providecommand{\newblock}{}

\end{document}